\documentclass[twocolumn]{aastex631}

\newcommand{\msolar}{\,$M_{\odot}$}
\newcommand{\mj}{\,$M_\mathrm{Jup}$}

\newcommand{\au}{\,au}
\newcommand{\um}{\,$\mu$m}
\newcommand{\muJy}{\,$\mu$Jy}
\newcommand{\wdA}{WD\,1202$-$232}
\newcommand{\wdB}{WD\,2105$-$82}
\newcommand{\teff}{${T}_{\mathrm{eff}}$}
\newcommand{\logg}{$\log{g}$}

\bibpunct{(}{)}{;}{a}{}{,}

\begin{document}

\title{JWST Directly Images Giant Planet Candidates Around Two Metal-Polluted White Dwarf Stars}

\author[0000-0001-7106-4683]{Susan E. Mullally}
\affiliation{Space Telescope Science Institute, 3700 San Martin Drive, Baltimore, MD 21218, USA}
\email{smullally@stsci.edu; n\'ee Susan E. Thompson}

\author[0000-0002-1783-8817]{John Debes}
\affiliation{AURA for the European Space Agency (ESA), Space Telescope Science Institute, 3700 San Martin Dr, Baltimore, MD 21218, USA}

\author[0000-0002-7698-3002]{Misty Cracraft}
\affiliation{Space Telescope Science Institute, 3700 San Martin Drive, Baltimore, MD 21218, USA}

\author[0009-0004-7656-2402]{Fergal Mullally}
\affiliation{Constellation, 1310 Point Street, Baltimore, MD 21231}

\author[0009-0008-7425-8609]{Sabrina Poulsen}
\affiliation{Homer L. Dodge Department of Physics and Astronomy, University of Oklahoma, 440 W. Brooks St, Norman, OK 73019, USA}

\author[0000-0003-0475-9375]{Loic Albert}
\affiliation{Institut Trottier de recherche sur les exoplanètes and Département de Physique, Université de Montréal, 1375 Avenue Thérèse-Lavoie-Roux, Montréal, QC, H2V 0B3, Canada}

\author[0009-0004-6806-1675]{Katherine Thibault}
\affiliation{Institut Trottier de recherche sur les exoplanètes and Département de Physique, Université de Montréal, 1375 Avenue Thérèse-Lavoie-Roux, Montréal, QC, H2V 0B3, Canada}

\author[0000-0001-8362-4094]{William T. Reach}
\affiliation{Stratospheric Observatory for Infrared Astronomy, Universities Space Research Association, NASA Ames Research Center, Moffett Field, CA 94035}

\author[0000-0001-5941-2286]{J. J. Hermes}
\affiliation{Department of Astronomy, Boston University, 725 Commonwealth Avenue, Boston, MA 02215, USA}

\author[0000-0001-7139-2724]{Thomas Barclay}
\affiliation{NASA Goddard Space Flight Center, 8800 Greenbelt Road, Greenbelt, MD 20771, USA}

\author[0000-0001-6098-2235]{Mukremin Kilic}
\affiliation{Homer L. Dodge Department of Physics and Astronomy, University of Oklahoma, 440 W. Brooks St, Norman, OK 73019, USA}

\author[0000-0003-1309-2904]{Elisa V. Quintana}
\affiliation{NASA Goddard Space Flight Center, 8800 Greenbelt Road, Greenbelt, MD 20771, USA}

\begin{abstract}

We report the discovery of two directly imaged, giant planet candidates orbiting the metal-rich DAZ white dwarfs \wdA\ and \wdB. JWST's Mid-Infrared Instrument (MIRI) data on these two stars show a nearby resolved source at a projected separation of 11.47 and 34.62\,au, respectively. Assuming the planets formed at the same time as their host stars, with total ages of 5.3 and 1.6 Gyr, the MIRI photometry is consistent with giant planets with masses $\approx$1--7\mj. The probability of both candidates being false positives due to red background sources is approximately 1 in 3000. If confirmed, these would be the first directly imaged planets that are similar in both age and separation to the giant planets in our own solar system, and they would demonstrate that widely separated giant planets like Jupiter survive stellar evolution. Giant planet perturbers are widely used to explain the tidal disruption of asteroids around metal-polluted white dwarfs. Confirmation of these two planet candidates with future MIRI imaging would provide evidence that directly links giant planets to metal pollution in white dwarf stars. 


\end{abstract}





\section{Introduction} \label{sec:intro}

While some of the first exoplanets ever discovered were found orbiting evolved stars \citep{Wolszczan1992Natur}, only a small fraction of the more than 5000 confirmed exoplanets are known to orbit non-main sequence stars. As a result, we have few observational constraints on what happens to planetary systems in the late stages of stellar evolution.  Most stars, and all low-mass stars, end their lives as white dwarf stars \citep[WDs,][]{fontaine01}. Finding exoplanets around these remnants can teach us about the fate of the planets in our own solar system.  

Theory suggests that exoplanets should exist around WDs. Outer planets (those in orbits beyond approximately the asteroid belt) should survive unscathed while planets inward of $\sim$1\,\au\ should be engulfed or tidally disrupted during the red giant phase \citep{Livio1984evolution, mustill2012}. Those that survive are expected to migrate outward due to the mass loss of the star.  
 Radial velocity surveys of evolved giant-branch stars currently undergoing mass-loss have found no difference in planet occurrence rates for giant planets compared to main-sequence stars \citep{Wolthoff2022}.

Despite several surveys of WDs using various techniques to find surviving massive giant planets \citep[e.g.][]{Burleigh2002, Debes2005II, Farihi08, Burleigh2008, Hogan2009,vansluijs2018, Brander2020}, only a few planetary mass objects have been found orbiting WDs. Recent examples of giant planets found orbiting single WDs include one found via micro-lensing \citep[MOA-2010-BLG-477Lb,][]{Blackman21} and another found transiting with a short orbital-period \citep[WD 1856+534 b,][]{Vanderburg2020Nature}. Also, \cite{Luhman2012}, using Spitzer IRAC, succeeded in finding a cool, low-mass, brown dwarf at a separation of 2500\,au with common proper motion to a WD.

Substantial evidence for planets around WDs comes from the presence of photospheric metals in 25\% -- 50\% of isolated, hydrogen-atmosphere WDs \citep{Koester2014}. These metal rich WDs must be actively accreting, since the strong gravitational field pulls heavier elements out of the atmosphere on timescales as short as a few days \citep{Koester2009diffusion} leaving a chemically pure exterior of hydrogen. Relic planetary systems are the favored theory for the source of the accreted material \citep{Alcock86, Jura03}. In this scenario, planets that survive the red-giant phase occasionally perturb the orbits of asteroids and comets \citep{DebesWalsh2012}, which then fall in towards the WD. When these bodies pass inside the Roche limit of the star they disintegrate into a cloud of dust and gas, which then accretes onto the star. 
Support for this theory comes from the similarity between the chemical composition of the accreted material and the composition of the bulk Earth \citep[e.g.][]{Melis2011, Xu2019, Trierweiler2023}.

JWST's infrared capabilities offer a unique opportunity to directly image Jupiter mass planets orbiting nearby WDs. The contrast between a cool planet and the small, hot WD is favorable, often better than 1:200.  Taking advantage of JWST's superb resolution, it is possible to directly image a planet at only a few au from nearby WDs without the use of a coronagraph. Additionally, because nearby WDs are typically several billion years old, planets found would be similar in age to those in our own solar system. A recent paper by \citet[][submitted]{Poulsen2023} demonstrates the exoplanet mass limits that can be achieved with JWST's MIRI imaging of both unresolved and resolved companions. Those observations are the deepest search to date for a WD; the 15\,\um\ band would have been able to detect a 3\,Gyr, 0.34\mj\ planet. That mass limit is $\approx$ 20 times better than achieved with previous Spitzer or Hubble observations of WDs \citep{Mullally2007Spitzer, Debes2005}.

In this paper we report evidence of directly imaged giant planets orbiting two nearby, metal-polluted WDs. Multi-band mid-infrared imaging of each WD shows evidence of a red point source around each target that is consistent with a giant planet.  We discuss potential sources of false positives and describe the observations required to confirm the candidates.  We conclude by discussing the implications of this discovery in regards to validating the theory that metal pollution on WDs are caused by giant planets. 


\section{Methods}
\subsection{White Dwarf Star Properties}
\wdA\,(LP 852-7) and \wdB\,(GJ820.1), were two of four stars chosen for deep MIRI observations as part of the GO program 1911 \citep{Mullally2021jwst.prop.1911M}. The four target stars (the other two were WD\,1620-391 and WD\,2149+02) were chosen to be isolated, to have metals in their atmospheres \citep[][]{Subsavage2017}, and to be either young or nearby to improve MIRI's sensitivity to giant exoplanets. Assuming any planet orbiting the WD formed at the same time as the star, younger stars will have warmer, brighter planets in the mid-infrared. Both stars have hydrogen atmospheres (type DA) and their metal lines indicate that material is actively accreting onto their surfaces. \wdB\, is also known to be magnetic with a polar field of roughly 50\,kG \citep{landstreet12}.

Table~\ref{tab:wdproperites} provides recent measurements of effective temperature, gravity, distance and brightness for each star.  This information is used to derive the age of the star, a key piece of information to determine the expected brightness of any planet found in the system.  The total age is made up of the cooling age plus the main sequence age. We calculate the total age using the wdwarfdate \citep{Kiman2022} software, and rely on the cooling models of \citet{Bedard2020}, the initial-to-final mass relation of \citet{Cummings2018}, and the stellar evolutionary models of \citet{Choi2016}. \wdA\ has an initial mass of 1.3 \msolar\ and a cooling age of 0.90\,Gyr, resulting in a total age of 5.3$^{+5.0}_{-2.5}$ Gyr. \wdB\ has an initial mass of 2.5 \msolar\ and a cooling age of 0.83\,Gyr, resulting in a total age of 1.6$^{+0.8}_{-0.2}$\,Gyr. The relatively large errors in the age measurement for \wdA\ is 
due to relatively large systematic uncertainties for the progenitor mass estimates for WDs with masses below 0.63\,\msolar\ \citep{heintz22}.

\begin{table*}[]
    \centering
    \begin{tabular}{c|llllllll}
    \hline
    \hline
  Name & $K_{\mathrm{mag}}$ &  \teff\ & \logg\ & Dist. &   $M_{\mathrm{WD}}$ & $M_\mathrm{{MS}}$ & Total Age \\
 & (Vega) & (K) & (cgs)  & (pc)  & (M$_{\odot}$) & (M$_{\odot}$) & (Gyr) \\ [0.5ex]
    \hline
         \wdA\   & 12.3  & 8760$\pm130$  & 8.01$\pm0.05$ &  10.43   & 0.60$^{+0.03}_{-0.02}$ & 1.3$^{+0.4}_{-0.3}$ & 5.3$^{+5.0}_{-2.5}$ \\
         \wdB\ &  13.5  & 9890$\pm170$ & 8.22$\pm0.08$ & 16.18  & 0.70$^{+0.06}_{-0.05}$ & 2.5$^{+0.6}_{-0.7}$ & 1.6$^{+0.8}_{-0.2}$ \\
    \end{tabular}
    \caption{Atmospheric parameters for \wdA\ based on the optical spectroscopy of \citet{Gianninas2011ApJ} with the 3D corrections of \citet{Tremblay2013}; for \wdB\ based on the radiative-atmosphere, optical spectroscopic fits of \citet{Gentile2018}. Ages and masses inferred using wdwarfdate \citep{Kiman2022}; distances from \citet{GaiaDR3}. }
    \label{tab:wdproperites}
\end{table*}

\subsection{MIRI Imaging} \label{sec:}
JWST targeted the WDs with the mid-infrared instrument (MIRI) in imaging mode with four different broadband filters: F560W, F770W, F1500W, F2100W.  The observations' exposures were designed to make it possible to detect at least a 1\mj\ planet orbiting the WD.  The observations of \wdA\ were taken on February 09, 2023 and the observations of \wdB\ were taken on April 21, 2023. Each image was composed of many exposures dithered using the cycling dither pattern with the FASTR1 readout pattern. The exposure times for the four filters for \wdA\ from shortest to longest wavelength were 255.3, 277.5, 8413.9 and 1309.8\,s. For \wdB\ the four filter exposure times were 233.1, 233.1, 12088.1, and 6016.3. 

The images were processed using build 9.2 (data processing software version 2022\_5c and calibration software version of 1.11.4) of the JWST Calibration pipeline 
\footnote{\url{https://jwst-docs.stsci.edu/jwst-science-calibration-pipeline-overview/jwst-operations-pipeline-build-information/jwst-operations-pipeline-build-9-2-release-notes}} starting from the uncal files, using a CRDS context from jwst\_1130.pmap and CRDS version of 11.17.6. Each set of images was processed through stage one and two of the imaging pipeline using mostly default parameters. After the stage two pipeline was run, a mean background image was created for each filter and subtracted from each cal file. Lastly, the level three imaging pipeline was run on the background subtracted files and combined with the resample kernel set to `gaussian' and the outlier detection `scale' values being set to (1.0 and 0.8), but using the standard parameter reference files for all other parameter settings. These background subtracted cal and i2d files were used for all subsequent analyses. All data can be found at \dataset[DOI: 10.17909/kak5-tx90]{http://dx.doi.org/10.17909/kak5-tx90}

    \label{tab:wdobsparams}


\begin{figure*}
    \centering
    \includegraphics[scale=0.45]{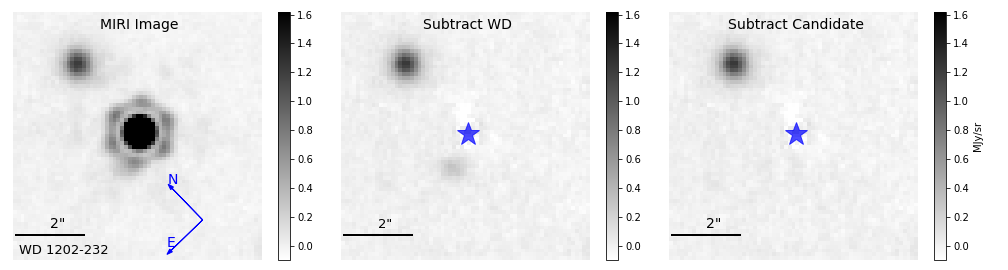}
    \includegraphics[scale=0.45]{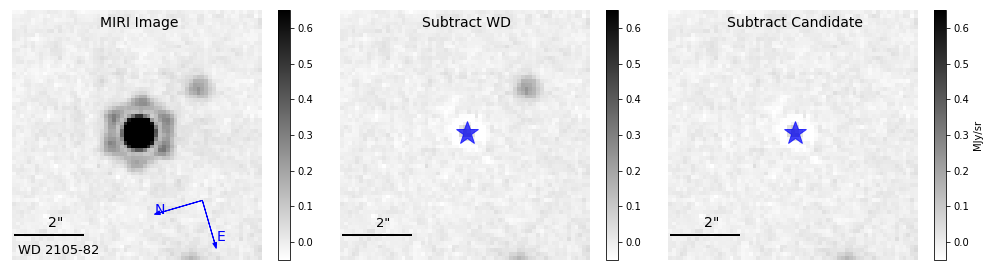}
    \caption{Image of each star and its candidate before and after subtracting a PSF modeled from a nearby bright star. Images cutouts are 65x65 pixels and the flux density units are in MJy per steradian. Left: Original calibrated MIRI image cutout centered on the WD. The north and east directions are specified with blue arrows.  Middle: The same image after scaling and subtracting the PSF centered on the WD. The location of the removed star is shown as a blue star. Right: The MIRI image after subtracting both the WD and the candidate using the same PSF. In both cases the candidate is removed cleanly, indicating it is point-source in nature. Note that the bright object north of \wdA\ is a galaxy. }
    \label{fig:psf}
\end{figure*}

\subsection{PSF Subtraction and Aperture Photometry of the Candidates}
To search for close companions to our targets, we performed reference differential imaging. After some experimentation, we found that the best reference PSFs for our program came from the other WDs--good PSF subtraction with MIRI imaging seems to require both a similar color as the target and similar placement on the detector. This is most noticeable for F560W and F770W, where we also need a good removal of the cruciform feature. We judge the best PSF in each case to be the one that leaves the smallest visible residuals and is hitting the noise floor as close as possible to the location of the star. We only found a point source near (within $\approx$3 arc seconds) of the target star for \wdA\ and \wdB. The other two stars show no evidence of a candidate \citep[for WD2149+021 see][]{Poulsen2023}. 

For \wdA, we constructed average PSFs for each filter and masked out background sources.  For F560W we found best results with a combination of WD~1620-391 and WD~2149+021 PSFs \citep{Mullally2021jwst.prop.1911M, Poulsen2023}. At F770W, only WD~1620-391 provided the best subtraction, and for F1500W and F2100W a combination of the PSFs for WD~1620-391, WD~2149+021, and \wdB
were satisfactory.  We used the same procedure for \wdB. In this case we made use of the PSFs for the other three WDs since the companion is well separated, however we obtained a better subtraction for F1500W by excluding \wdA. Even without PSF subtraction, the two candidates are easily discernible, and the PSF subtraction does not reveal any other candidates closer than those detected. 

In order to determine the flux of the candidates in each filter, we performed aperture photometry. First, to get a high quality absolute flux measurement of the star in each filter, we used aperture photometry on the target star first using the techniques described by \citet{Poulsen2023}. To maximize signal-to-noise on the candidates, we then calculated the photometry in an aperture equivalent to half the PSF full-width half max in each filter and performed the same measurement on the unsubtracted star's PSF. We then took the ratio between the candidate and star's photometry calculated using this small aperture. We then multiplied this ratio by the star's absolute flux measurement in each filter. The extracted photometry for each filter are reported in Table~\ref{tab:phot}.


\section{Results}

The goal of this study is to identify candidate substellar companions to the WDs. We expect these to appear as point sources that increase in brightness at longer wavelengths.  The expected shape of the spectral energy distribution of old, cool giant planets in these broad band filters is not well known; models show that temperature, composition and cloud-cover can all impact these bands.  Here we discuss the evidence for directly imaged planet candidates in the MIRI observations. We compare the four band spectral energy distribution (SEDs) of both candidates with planet models and constrain the mass of the planet assuming the observed sources lie at the distance of the star and have an age matching that of the WD. We also examine whether the target is a point source by looking at the Gaussian shape of the center of the source and by subtracting a PSF from the companion and looking at the residuals. Finally we explore the overall likelihood of a false positive.


\subsection{Planet Candidates}

\paragraph{\wdA\,b}  Figure~\ref{fig:psf} shows the MIRI $15\mu$m image of \wdA\ (top panels) before and after PSF subtraction of the WD and a nearby source, which is clearly visible at a separation of 1.11$\pm0.04$ arc seconds at a position angle of 114$\pm$2 degrees from North. These position measurements come from averaging the separation measured independently in each of the four filters. The absence of significant residuals in the PSF-subtracted images indicate the nearby source is not extended; all remaining flux is consistent with residuals after subtracting the star.
Given a distance of 10.43~pc for \wdA, if the planet is in a circular, face-on orbit, the star-planet separation is 11.47\,\au\ and the orbital period is 50 years. During the host star's main-sequence lifetime, the planet's orbit would have been 5.3\,\au, assuming the planet moved out adiabatically after the host-star mass loss phase. 

\begin{figure*}
\centering
\includegraphics[scale=0.27]{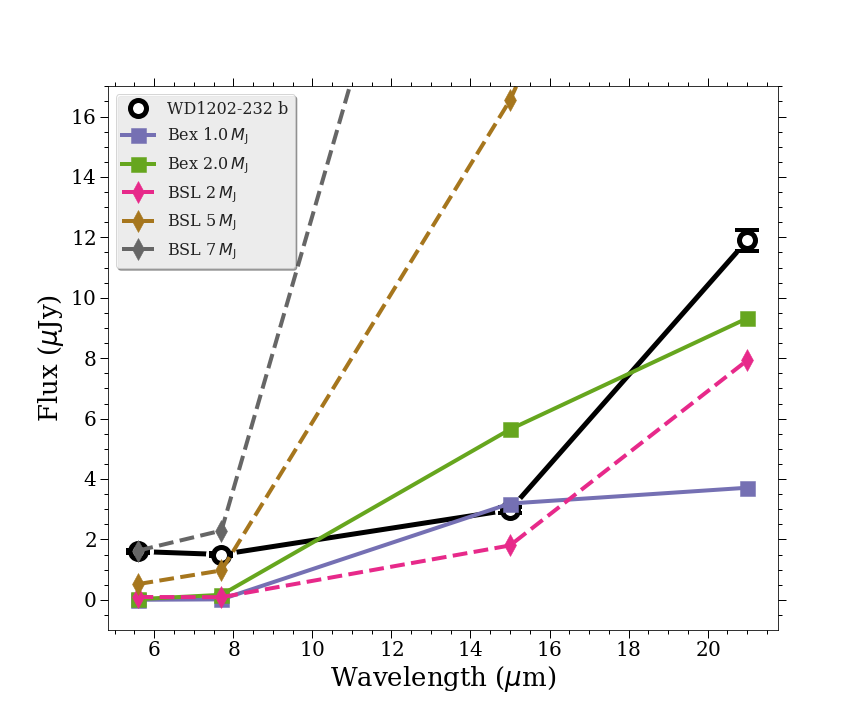}
\includegraphics[scale=0.27]{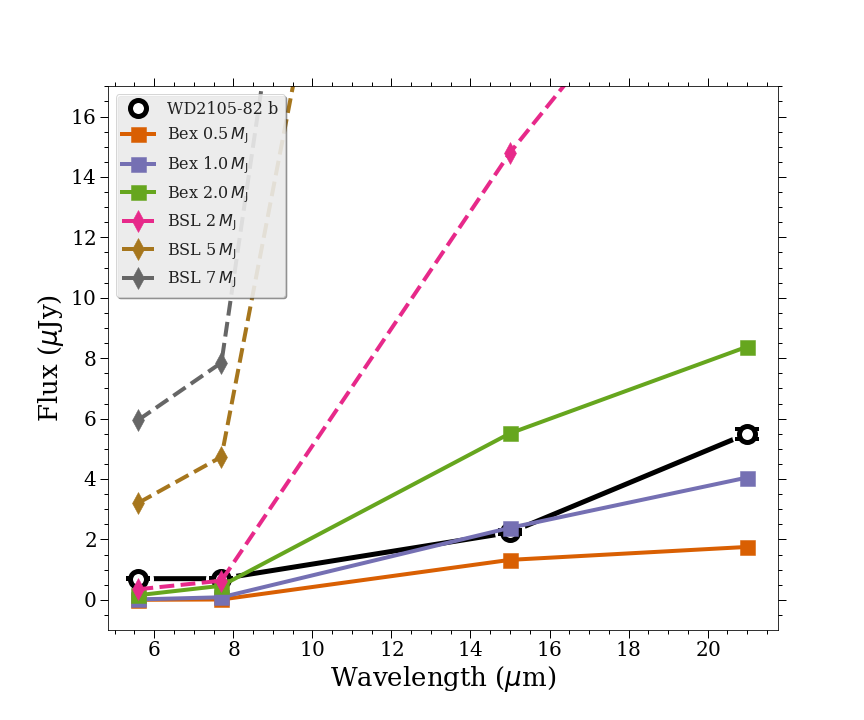}
\caption{Photometric measurements for all four bands for both planet candidates (black circles) compared to low mass planet models  (squares) using the BEX \citep{Linder2019} and BSL \citep{Burrows2003} models.  \wdA\,b is shown in the left compared to models of 5.25\,Gyr and \wdB\,b is shown in the right compared to models at 1.6\,Gyr for BEX and 1.0\,Gyr for BSL (due to the available grid). While the photometry does not exactly match the models, the flux in all four bands is consistent with giant-mass planet at the age of the WD and shows increasing flux with wavelength.}
\label{fig:sed}
\end{figure*}

The candidate has a 15\,\um\ flux of 3.0$\pm$0.1\muJy, consistent with a cool planetary-mass object at the distance of the star. The same source is seen in all four bands. Figure~\ref{fig:sed} shows the SED of the candidate planet around \wdA\ (left panel)
along with the predicted SEDs from the Bern EXoplanet Cooling Curves with solar metallicity and no clouds \citep[BEX,][]{Linder2019} for planet masses less than 2\mj\ and the \citet{Burrows2003} models for higher mass planets.  Using the the total age of the star as the cooling time for the planet, this point source is consistent with the expected flux for a planetary mass object of approximately 1 -- 7\mj. 

\paragraph{\wdB\,b} The second candidate orbits \wdB. MIRI imaging revealed a mid-infrared point source well separated from the WD at a separation of 2.14$\pm$0.02 arc seconds at a position angle of 200.4$\pm$0.4 degrees from North.  Given the distance to the WD, if the candidate is in a face-on circular orbit, this translates to a star-planet separation of 34.62\,au and a 243 year Keplerian orbit. During the main-sequence phase, the planet would have been at 9.7\,au. Figure~\ref{fig:psf} (lower panels) shows the candidate is a point source that is well subtracted in all filters using a PSF model created using a nearby bright star. A comparison with the BEX models \citep{Linder19} shown in the right panel of Figure~\ref{fig:sed} indicates that the candidate planet's mass ranges from approximately 1 to 2\mj.  In this case the age of the WD is relatively well constrained, and the error in the planet mass is dominated by the photometric errors.

For both \wdA\ and \wdB, the model spectral energy distributions (SEDs) in the four MIRI bands do not exactly match the observations. In general, the 5.6 and $7.7\,\mu$m bands are brighter than expected compared to the 15 and $21\,\mu$m  bands. However, this is not sufficient evidence to rule-out these candidates as planets. Planet models at these masses and temperatures are relatively unconstrained and they depend on the metallicity of the planet as well as the presence of clouds and haze in their atmospheres \citep{Marley2021, Limbach2022}. These unknowns can have significant impact on the shape of the exoplanet's SED. Other, more exotic effects, such as tidally heated exomoons \citep{limbach13}, could also cause the models to not match. 

\begin{table}
    \centering
    \begin{tabular}{l|llll}
    \hline
    \hline
    Object & F560W & F770W & F1500W & F2100W \\
       & ($\mu$Jy)   & ($\mu$Jy) & ($\mu$Jy) & ($\mu$Jy) \\
     \hline
    \wdA\ & 1208$\pm$36&699$\pm$21 &181$\pm$5 & 105$\pm$3 \\
    \wdA\,b & 1.6$\pm$0.1 &1.5$\pm$0.1 & 3.0$\pm$0.1 & 11.9$\pm$0.6\\
    \wdB\ &428$\pm$13 &229$\pm$7 & 68$\pm$2 & 35$\pm$1\\
    \wdB\,b & 0.7$\pm$0.2&0.7$\pm$0.2 & 2.2$\pm$0.1 & 5.5$\pm$0.3\\
    \end{tabular}
    \caption{Photometric measurements in the 4 JWST bands of the WDs and the candidate companion to the WDs. Error bars all include a 3\%\ error bar from photometric calibrations. This is the dominant error for the WD photometry.}
    \label{tab:phot}
\end{table}

\subsection{Potential False Positives}
While the available observations are consistent with an orbiting exoplanet, we also consider other objects that would be consistent with the available evidence.  The SED is inconsistent with a chance alignment with a typical main sequence or evolved red giant star because even the coolest of these stars ($\sim$2000\,K) would be brighter at 5\um\ than at 15\um\, \citep{phoenix2013} unless enshrouded in dust. A brown dwarf is possible, but if so, the flux of our candidates would place it farther away and unbound to the WD. Given the scarcity of isolated brown dwarf stars, this scenario is unlikely. An object in our own solar system, such as a trans-Neptunian object of the appropriate size, could appear as a very red point source, but likely would move by several pixels during the exposures. Additionally both WDs lie at modest to high ecliptic latitudes ($-21^{\circ}$ and $-60^{\circ}$) where the density of such objects is low.

The most likely false positive scenario is a distant galaxy. Galaxies have a variety of colors and can appear as a point source if sufficiently small or far away. While recent galaxy count studies with JWST can give a worst case scenario to these odds \citep{Ling2022, WuCossas23}, we can get a more precise feel for the likelihood of a false positive by counting the number of background sources in all four MIRI fields (9237 square arc seconds per field) and determining the density of dim, red, point sources.

Following the same procedures outlined in \citet{Poulsen2023} we performed aperture photometry using the recommended aperture radii and background apertures across the entire field of view for both sets of observations. The core of each source was fit with a Gaussian to determine if its shape was consistent with a point source. We count red sources whose F1500W/F770W and F2100W/F1500W flux ratios are both greater than one, as expected for a potential companion. We also limit the search to sources whose flux in the F1500W filter is less than 100\muJy\ because the BSL models find all planets less than $\approx$13 Jupiter masses at these target's distances and ages would be dimmer than this value.
Counting across all four fields we find an average of 5 red point sources per MIRI field. The variation between fields range from 3 to 8 dim, red objects per field, as one would expect based on counting error. For all fields we use the average density of background sources to determine the rate of background sources, $0.54\times 10^{-3}$ sources per square arc second. 

The larger the area of the search, the more likely we will include a background false positive. Within a 1.1 arc second radius search (not including the inner 0.5 arc seconds we could not search because of the target star), we found one dim, red candidate in all four fields, \wdA~b. The odds of finding one object in four fields is inconsistent with the observed background density at a 2-sigma level at 2.8 arc seconds or smaller. Each candidate, if it were the only candidate found, has a false alarm rate better than a 2-sigma result, 1/128 for \wdA\ and 1/29 for \wdB. However, we actually found two candidates around four targets within 2.14 arc seconds. Using binomial statistics, finding two objects in four fields is inconsistent with the observed background density at a level of 1 in 3000 for a search radius of 2.2 arc seconds or smaller. We conclude that both are exoplanet candidates and that it is very unlikely that both will be shown to be false positives.

\newpage
\section{Discussion}

The best way to definitively confirm these candidate planets is to obtain a second epoch of the field with MIRI. The  proper motion of \wdA\ is 230\,mas/yr and of \wdB\ is 463 mas/yr \citep{GaiaDR3}. Because MIRI's pixels are 110 mas across \citep{jdox}, this translates to a proper motion of 2 pixels/year for \wdA\ and 3.6 pixels/year for \wdB. A second observation with JWST/MIRI could confirm common proper motion and the planetary nature of each candidate.  

Current models for what drives the pollution on these WDs indicate that many could have observable planetary companions.  \citet{Veras21} lists a variety of models that seek to explain the source of WD metal pollution. While all invoke the destruction of some kind of minor planet (a comet or asteroid) as a direct explanation of the pollution, models differ in how to transport those minor planets from beyond 3\au\ to the Roche limit where they are disrupted. The standard model of \citet{Debes2002}, \citet{Frewen14} and others invoke changes in the orbit of a large planet (either due to stellar mass-loss in the asymptotic giant branch stage, or planet-planet scattering) to drive orbital instability in the minor planets (the so-called very late heavy bombardment). If that theory is correct, we expect all such polluted stars to host a giant planet ($\gtrsim$ 0.1\mj, roughly twice Neptune's mass). While MIRI images are not sensitive to the smaller giant planets, it is likely to be sensitive to planets above $\approx$ 1\mj\ as shown by \citet{Poulsen2023} for similar observations. If the mass distribution for giant long-period planets reported by \citet{Fernandes19} is accurate, JWST observations should be sensitive to approximately a third of the expected giant companions. Failure to find any planets orbiting four WDs would start to shift the burden of evidence strongly in favor of models such as \citet{Caiazzo17} and \citet{Veras23} that invoke much smaller planets to explain the accretion of materials onto WDs. A future paper will discuss detection limits and any widely-separated candidates for the entire sample of MIRI observations.


If confirmed, these two planet candidates provide concrete observational evidence that outer giant planets like Jupiter survive the evolution of low mass stars. Their existence would support the dominant paradigm for the planet-pollution connection, arguing that wide-orbit giant planets are ubiquitous. Confirmation would also support the indirect evidence that 25--50\% of WDs host large planets \citep[as infered from the fraction of metal-polluted WDs,][]{Koester2014}.  This would mean that the occurrence rate of giants around WD progenitors (B--F type stars) is also high. Confirmation would support the conclusions of ground-based direct imaging surveys of young stars \citep{Nielsen2019GPI, Vigan2021} that B and A stars have a higher fraction of giant planets than solar type stars. The confirmation of these planets are not, however, sufficient to fully validate that large-mass giant planets are the driver of accretion without further observations. Only by surveying more nearby WDs, both with and without metal lines in the atmosphere, can we determine whether large giant planets are more common around metal polluted WDs. A first step towards such a survey will be undertaken by JWST starting in Cycle 2 through the MEAD (GO\,3964) and MEOW (GO\,4403) survey programs which may be sensitive to the brighter, more massive, giant planets.

\begin{figure}
    \centering
    \includegraphics[width=1.05\linewidth]{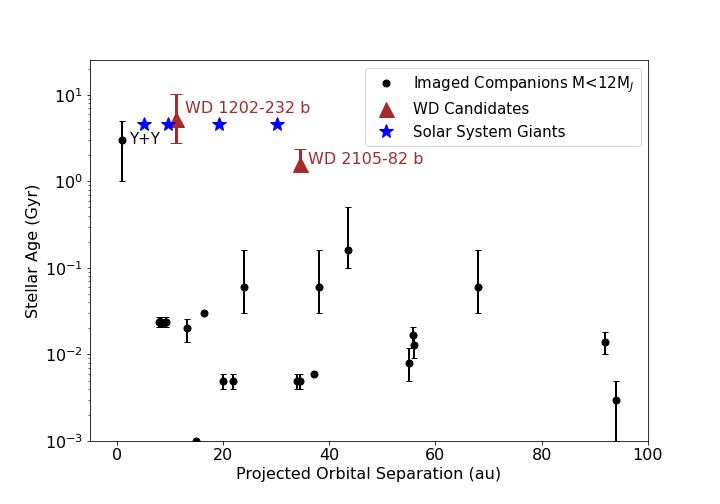}
    \caption{Projected orbital separation versus the age of the star the companion is orbiting for directly imaged planets less than 12\,\mj\ (black circles). The two WD planet candidates are presented with brown triangles. The solar system giant planets are shown as blue stars. The candidates are significantly older than all previously reported directly imaged, low mass companions, except the Y+Y dwarf binary system (WISE J033605.05-014350.4) recently found using JWST by \citet{Calissendorff2023}. }
    \label{fig:planets}
\end{figure}

These candidates would represent the oldest directly imaged planets outside our own solar system, and in many ways are more like the planets in our outer solar system than ever discovered before. Figure~\ref{fig:planets} shows a comparison with confirmed planets and brown dwarfs compiled by the NASA exoplanet archive \dataset[DOI: 10.26133/NEA12]{http://dx.doi.org/10.26133/NEA12}.  These candidate planets are likely only a few times the mass of Jupiter and their separation during the main sequence (5.3\,\au\ and 9.7\,\au) are similar to the orbits of Jupiter and Saturn in our own solar system.  Also, the ages of these planets, and hence their temperatures, are more similar to the 4.6\,billion year old age of our own solar system than previously directly imaged exoplanets around much younger stars \citet[e.g.][]{Nielsen2019GPI}. Thermal emission from a cool, middle-aged exoplanet provides critical constraints on atmospheric properties that also fold back into a greater understanding of the cold giant planets in our Solar System. Direct detection of the thermal emission from cold (\teff$<$200~K) planets is particularly important for understanding the presence or absence of water clouds, similar to that seen for Jupiter \citep{Marley2021, Morley2014}.  The impact of clouds on exoplanet emission could be significant; cloudy planets lose their prominent emission at 4.5\,\um\ while colder, cloudless planets may be fainter at wavelengths beyond 10\,\um\ \citep{Limbach2022}. 

\section{Conclusions}

We have found evidence of two giant planets orbiting two different DAZs using broad band mid-infrared imaging with JWST's MIRI. The sensitivity and resolution of MIRI along with the light gathering power of JWST, have made it possible to image previously unseen middle-aged giant planets orbiting nearby stars, all without a coronagraph. The most likely false positive scenario is a distant red galaxy, however background source counts indicate finding such a galaxy so close to both WDs at 15\,\um\ is highly unlikely. If confirmed using common proper motion, these giant planets will represent the first directly imaged planets that are similar in age, mass, and orbital separation as the giant planets in our own solar system.  Future spectroscopy and multi-band imaging of these systems may be possible with JWST, which would improve the observational constraints on the physics and variety of cool giant planet models.

\begin{acknowledgments}
The authors would like to thank the anonymous referee for providing very useful comments that improved this paper. This work is based on observations made with the NASA/ESA/CSA James Webb Space Telescope. The data were obtained from the Mikulski Archive for Space Telescopes at the Space Telescope Science Institute, which is operated by the Association of Universities for Research in Astronomy, Inc., under NASA contract NAS 5-03127 for JWST. Support for program number 1911 was provided through a grant from the STScI under NASA contract NAS5-03127. MK acknowledges support by the NSF under grant AST-2205736 and NASA under grant 80NSSC22K0479.

\end{acknowledgments}
\software{Astropy \citep{astropy}, Matplotlib\citep{matplotlib}}

\bibliography{dusty-wd}{}
\bibliographystyle{aasjournal}

\end{document}